\documentclass{aa}  
\usepackage{subfig}
\usepackage{graphicx}
\graphicspath{{figures/}}
\usepackage[varg]{txfonts}
\usepackage{commath}
\usepackage{lastpage}
\usepackage[super]{nth}
\usepackage{siunitx}[=v2]
\usepackage{float}
\newcommand{\floor}[1]{\left\lfloor #1 \right\rfloor}

\DeclareSIUnit{\erg}{erg}
\DeclareSIUnit{\jansky}{Jy}
\DeclareSIUnit{\parsec}{pc}

\usepackage[dvipsnames]{xcolor}
\definecolor{click_color}{RGB}{46,48,118}

\usepackage{hyperref}
\usepackage[capitalize]{cleveref}
\crefname{equation}{eq.}{eqs.}
\crefname{section}{Sect.}{Sects.}
\begin{document}

   \title{Chromatic activity window of periodic fast radio bursts: FRB~20121102A and FRB~20180916B }

   % \author{
   %        \inst{1} \and M. Cruces \inst{3} \and T. Cassanelli \inst{4} \and C. Braga \inst{2}
   %        }

    \author{M. C. Espinoza-Dupouy\thanks{Email: maria.espinoza.d@ug.uchile.cl}
          \inst{1}
          \and M. Cruces \inst{2,3,4,5,6}
          \and T. Cassanelli \inst{7} \and C.A. Braga \inst{2,8} \and E. Bermúdez \inst{2} 
            \and J. Vera-Casanova \inst{2,3} 
          }

    \institute{
    Department of Astronomy, Universidad de Chile, Camino El Observatorio 1515, Las Condes, Santiago, Chile
    \and
    Centre of Astro-Engineering, Pontificia Universidad Católica de Chile, Av. Vicuña Mackenna 4860, Santiago, Chile
    \and
     Department of Electrical Engineering, Pontificia Universidad Católica de Chile, Av. Vicuña Mackenna 4860, Santiago, Chile
    \and
    Joint ALMA Observatory, Alonso de Córdova 3107, Vitacura, Santiago, Chile
    \and
    European Southern Observatory, Alonso de Córdova 3107, Vitacura, Casilla 19001, Santiago de Chile, Chile
    \and
    Max-Planck-Institut für Radioastronomie, Auf dem Hügel 69, D-53121 Bonn, Germany.
    \and
    Department of Electrical Engineering, Universidad de Chile, Av. Tupper 2007, Santiago
    \and
     Instituto de Astrofísica, Facultad de Física, Pontificia Universidad Católica de Chile, Casilla 306, Santiago 22, Chile
    }

   \date{Received ?, 2025; accepted ?, 2025}

\abstract
{Two fast radio bursts, FRB~20121102A and FRB~20180916B, show periodic activity with cycles of \num{159.3} and \num{16.33}~days, respectively. These cycles consist of active and inactive windows, with the peak activity centred within the active phase. For FRB~20180916B, studies reported a frequency-dependent or chromatic behaviour, for which the activity window starts earlier and becomes narrower at higher frequencies. The activity across frequencies is typically modelled with a power law. 

}
{We developed a simple model that combines the phase and frequency dependence of the activity windows of FRB~20121102A and FRB~20180916B. Our goal was to perform a chromaticity study for FRB~20121102A that incorporates model improvements to account for the cyclic nature of its activity window and to compare the chromatic behaviour of the two periodic FRBs.

}
{We standardised the detections from the 425 observing epochs for FRB~20121102A and the 214 epochs for FRB~20180916B to account for differences in the radio telescope sensitivity. To the normalised detection-rate phase distribution, we fitted a von Mises distribution and extracted the peak activity phase and activity width. These quantities as a function of frequency were then modelled as power laws to construct the chromatic model.
}
{The activity window starts earlier at higher frequencies for the two sources. The activity window of FRB~20121102A broadens at higher frequencies, however, and that of FRB~20180916B broadens at lower frequencies. Interestingly, it appears to remain active during at least 80\% of the cycle at C band.
}
{The observed chromatic behaviour of FRB~20180916B is consistent with previous findings. For FRB~20121102A, a chromaticity in its activity window is also seen, but the source appears to be active for longer at higher frequencies, which is different from the behaviour of FRB~20180916B.
}


   \maketitle

\section{Introduction}

First discovered by \cite{lorimer2007bright}, fast radio bursts (FRBs) are an astronomical phenomenon that consists of exceptionally bright (\SIrange{e36}{e41}{\erg}, \citealt{zhang2018frb, locatelli}) coherent radio pulses with durations from micro- to milliseconds \citep{zhang2023physics}. In addition to the Galactic magnetar SGR~1935$+$2154, which was observed to emit millisecond radio pulses \citep{2020Natur.587...54C}, most FRBs are known to have an extragalactic origin. This is supported by their observed dispersive delay contribution, which is consistent with localised host galaxies \citep{chatterjee2017direct, marcote2022precise, tendulkar2017host, petroff2022fast, 2024NatAs...8.1429C}. These delays are quantified through the dispersion measure (DM; in \si{\parsec\per\centi\m\cubed}) relation, 
\begin{equation}
  \text{DM} = \int_0^L n_e(L^\prime) \, \dif{L^\prime},
\end{equation}
which refers to the number density of electrons ($n_e$) integrated along the path travelled by the photons from the source to the observer. The observed dispersive delay $(\Delta t)$ is expressed as:
\begin{equation}
    \Delta t = {k_\text{DM}} \times \left(\frac{1}{{\nu_\text{low}}^{2}} - \frac{1}{{\nu_\text{high}}^{2}} \right) \times \text{DM},
	\label{eq:dispersion}
\end{equation}
where ${k_\text{DM}} = 1/\del{\num{2.41e-4}}$ \si{\per\parsec\centi\m\cubed\mega\hertz\squared\s}, ${\nu_\text{low}}$ and ${\nu_\text{high}}$ are the lower and higher passband frequencies (\si{\mega\hertz}) (\citet{handbook}).

The discovery of the first repeating FRB, FRB~20121102A \citep{spitler2016repeating}, led to the classification of FRBs into two categories: repeaters and one-offs. To date, around \num{\sim800} FRB sources have been reported, and only \SI{8.6}{\percent} were identified as repeaters \citep{Xu_2023}. According to \citet{Law_2022}, as few as \SI{1}{\percent} are highly active repeaters. This subsample corresponds to the sources with a high potential for multi-wavelength follow-up and long-term monitoring.

Three objects are reported to have periodic activity, FRB~20121102A \citep{Rajwade_2020, cruces2021repeating, Braga_2025}, FRB~20180916B \citep{2020Natur.582..351C}, and FRB~20240209A \citep{2025arXiv250211215P}, which have alternating on- and off-activity cycles. The period of FRB~20121102A is approximately \num{159} days, with an active phase lasting \SI{53}{\percent} of this cycle \citep{Braga_2025}. FRB~20180916B displays a \num{\sim16}-day periodicity and an on-phase of about \SI{31}{\percent} \citep{2020Natur.582..351C, bethapudi2023high}. Most recently, it was reported that FRB~20240209A follows a \SI{\sim126}{-day} cycle, with an activity window covering nearly \SI{70}{\percent} of the period \citep{2025arXiv250211215P}. These cases might indicate that multiple repeating FRBs have an active window that repeats at different period lengths, and some are significantly larger. This might explain the lack of a detected periodicity for some repeating FRBs.

As noted by \citet{pastor2021chromatic}, the activity window of FRBs seems to depend on the frequency. These authors studied FRB~20180916B from \SIrange{140}{1400}{\mega\hertz} and observed that the active window begins earlier and becomes narrower at higher frequencies. This phenomenon is referred to as chromaticity. Studying FRB~20180916B, \citet{bethapudi2023high} later extended this finding using a non-cyclic power-law model that relates detected burst with the observing frequency. They detected bursts at higher frequencies (\SIrange{4}{8}{\giga\hertz}), which supported the behaviour of the activity window described by \citet{pastor2021chromatic}.

To date, the progenitors of FRBs remain unknown. The periodic and chromatic behaviour observed in this subset of repeating FRBs enable targeted follow-up and multi-wavelength campaigns, however. These observations are crucial for probing their emission mechanisms and local environments, and they thereby help us to constrain possible progenitor models.

We model the activity window of periodic FRBs as a function of frequency ($\nu$) and phase ($\phi$). In Sect. \ref{sec:data_compi} we describe the  compilation and combination of all the publicly available observations. In Sect. \ref{sec:model} we introduce the model and apply it to the active windows. In Sect. \ref{sec:results} we show the resulting chromatic active windows and forecast activity cycles. In Sect. \ref{sec:disc} we discuss the behaviour of the active windows for FRB~20121102A and FRB~20180916B, highlight differences between them, and address the limitations of the model. We also examine the relation of the burst energy distribution to the active window structure. Finally, in Sect. \ref{sec:conclusions}, we summarize our main findings.

\section{Data compilation and combination}
\label{sec:data_compi}

We have compiled publicly available data from FRB~20121102A and FRB~20180916B. In particular, we were interested in the start of the observation (modified Julian date; MJD) and its duration, the detection count, the central frequency, and the observing bandwidth. When only the burst MJDs were reported, we used the time of arrival (ToA) between the first and last detected burst to estimate the observation duration.

The dataset included observations made by the following telescopes: Effelsberg (EFF), the Arecibo Observatory (AO), the Westerbork Synthesis Radio Telescope (WSRT{1131049}), the Five-hundred-meter Aperture Spherical radio Telescope (FAST), the Green Bank Telescope (GBT), the Lovell Telescope, the Meerkat Telescope, the Very Large Array (VLA), Deep Space Network telescopes 43 and 63 (DSS-43 and DSS-63), the Canadian Hydrogen Intensity Mapping Experiment and its FRB backend (CHIME/FRB), the Low-Frequency Array (LOFAR) and the Giant Metrewave Radio Telescope (GMRT).

As there is evidence of chromaticity in the activity window (\citealt{bethapudi2023high, pastor2021chromatic}), we labelled the datasets with their frequency bands: The P band covers from \SI{100}{\mega\hertz} to just below \SI{1}{\giga\hertz}, the L band covers frequencies from \SI{1}{\giga\hertz} to just below \SI{2}{\giga\hertz}, the S band spans from \SI{2}{\giga\hertz} to just below \SI{4}{\giga\hertz}, and the C band ranges from \SI{4}{\giga\hertz} to \SI{8}{\giga\hertz}. \Cref{tab:observational-data} presents the overview of all of the collected data for each telescope, and Sect. \ref{sec:model} presents the bins we used to discretise the frequency bands further.

\begin{table*}[t]
\caption{\label{tab:observational-data}
Observational campaigns for FRB~20121102A and FRB~20180916B.}
\centering
\begin{tabular}{lccc}
\hline\hline
\multicolumn{4}{c}{\textbf{FRB~20121102A}}\\
\hline
Name &
$F_{\mathrm{min}}$ (\si{\jansky\milli\s}) &
Frequency band &
Total observations \\
\hline
EFF        & 0.15 / 0.12              & L-band / C-band                    & 70  \\
Arecibo   & 0.03 / 0.02 / 0.01        & L-band (ALFA / L-Wide) / C-band     & 118 \\
GBT       & 0.05 / 0.02               & S-band / C-band                    & 29  \\
FAST      & 0.012                     & L-band                             & 68  \\
WSRT      & 0.26                      & L-band                             & 11  \\
MeerKAT   & 0.06                      & L-band                             & 1   \\
DSS-43    & 0.23                      & S-band                             & 1   \\
DSS-63    & 0.50                      & S-band                             & 18  \\
VLA       & 0.19 / 0.06 / 0.03        & L-band / S-band / C-band           & 47  \\
Lovell    & 0.23                      & L-band                             & 87  \\
CHIME/FRB & 0.352                     & P-band                             & 1   \\
\hline
\multicolumn{4}{c}{\textbf{FRB~20180916B}}\\
\hline
Name &
$F_{\mathrm{min}}$ (\si{\jansky\milli\s}) &
Frequency band &
Total observations \\
\hline
EFF        & 0.0875 & C-band & 30 \\
CHIME/FRB  & 0.352  & P-band & 35 \\
LOFAR     & 0.748  & P-band & 58 \\
GBT        & 0.110  & P-band & 9  \\
GMRT       & 0.643  & P-band & 15 \\
Apertif   & 0.406  & L-band & 52 \\
uGMRT     & 0.719  & L-band & 23 \\
\hline
\end{tabular}

\tablefoot{Summary of the observational campaigns for FRB~20121102A and FRB~20180916B. The table lists the telescopes used, their minimum fluence thresholds ($F_\text{min}$), the frequency bands of observation, and the total number of observations.}
\end{table*}

\subsection{Dataset for FRB~20121102A}

The FRB~20121102A dataset spans from November \num{2}, 2012, to November \num{17}, 2023. These \num{\sim11} years of data consist of \num{153} detections and \num{272} non-detections over \num{425} epochs.

The P-band dataset uses the observations reported by \citet{josephy2019chime}.

The L-band dataset contains observations reported by  \citet{2014ApJ...790..101S}, \citet{2016ApJ...833..177S}, \citet{2016Natur.531..202S}, \citet{2017ApJ...846...80S}, \citet{2017MNRAS.472.2800H}, \citet{2017ApJ...834L...8M}, \citet{2019A&A...623A..42H}, \citet{2020A&A...635A..61O}, \citet{2020MNRAS.495.3551R}, \citet{2020MNRAS.496.4565C},
\citet{2021Natur.598..267L}, \citet{2021MNRAS.500..448C}, \citet{2022MNRAS.515.3577H}, \citet{2023MNRAS.519..666J}, \citet{2023ATel15980....1F}, \citet{2024SciBu..69.1020Z} and \citet{Braga_2025}.

The S-band dataset observations were reported by \citet{2016ApJ...833..177S}, \citet{2017ApJ...850...76L}, \citet{2017ApJ...846...80S}, \citet{2020ApJ...897L...4M}, \citet{2020ApJ...905L..27P}, and \citet{2021MNRAS.500..448C}.

Finally, the C-band observations were reported by \citet{2016ApJ...833..177S}, \citet{2017ApJ...850...76L}, \citet{2018ApJ...863..150S}, \citet{2018Natur.553..182M}, \citet{2018ApJ...863....2G}, \citet{zhang2018frb}, \citet{2021MNRAS.500..448C} and \citet{hilmarsson2021rotation}.

\subsection{Dataset for FRB~20180916B}

The data collection for FRB~20180916B spanned a period of 3 years, starting on September \num{16}, 2018, until December \num{1}, 2021. There are \num{132} detections and \num{82} non-detections over a span of \num{214} epochs.

The P-band dataset contains observations reported by  \citet{2020Natur.582..351C}, \citet{Pleunis_2021}, \citet{Sand_2022}, \citet{pastor2021chromatic}, and \citet{Bethapudi_2025}. The L-band dataset contains observations reported by  \citet{pastor2021chromatic}, and \citet{Bethapudi_2025}. The C-band dataset contains observations reported by \citet{bethapudi2023high}

\subsection{Combination of datasets}

To account for observational bias in telescope sensitivity, we chose to scale the detection rates as they would have been seen by a 100 m radio telescope with properties similar to that of the Effelsberg telescope. We used a simple scaling relation following
\begin{equation}
    R_\mathrm{100m} = R_\mathrm{i} \left( \frac{F_\mathrm{100m}}{F_\mathrm{i}} \right) ^{\gamma},\label{eq:rates}
\end{equation}
where $R_\mathrm{i}$ stands for the observed rate (events detected per unit time), $F_\mathrm{i}$ and $F_\mathrm{100m}$ are the fluence thresholds of the telescope in which the observations were made and the 100 m telescope of reference, respectively, both calculated by  $F = \mathrm{S}_\text{min} \times\Delta t$, where $\mathrm{S}_\text{min}$ is the minimum detectable spectral flux density and $\Delta t$ is the pulse width. The exponent \( \gamma \) reflects the scaling \( R \propto E^{\gamma} \) between the rates and the energy, which varies for each FRB. For FRB~20121102A, this factor is between $\gamma = \num{-1.1}$ and $\gamma = \num{-1.8}$ \citep{cruces2021repeating, gourdji2019sample}, and we used $\gamma = \num{-1.1}$ based on \cite{cruces2021repeating}. For FRB~20180916B, the factor ranges from $\gamma = \num{-1.4}$ to $\gamma = \num{-0.5}$  \citep{pastor2021chromatic, bethapudi2023high}, and we chose $\gamma = \num{-1.3}$ following \citep{2020Natur.582..351C}.
To compute the minimum spectral flux density $\mathrm{S}_\text{min}$, we used the radiometer equation,
\begin{equation}
  \mathrm{S}_\text{min} = \frac{\mathrm{S/N}_\text{min} \times \mathrm{SEFD}}{\sqrt{n_{\mathrm{p}} \times BW\times \Delta t}}, \label{eq:minflux}
\end{equation}
where $\mathrm{S/N}_\text{min}$ is the minimum signal-to-noise ratio, SEFD is the system equivalent flux density of a telescope, $n_{\mathrm{p}}$ is the number of polarizations, and $BW$ is the receiver bandwidth.

%a power law index
The reference parameters for the 100 m telescope include an SEFD of \SI{15}{\jansky}, a minimum detectable $\mathrm{S/N}_\text{min}$ of \num{7}, a pulse width of $\Delta t =  \SI{1}{\milli\s}$, and two polarisations.  

The advantage of Eq. \ref{eq:rates} lies in its simplicity, which offers a straightforward scaling relation while assuming a representative telescope size that falls within the middle range of instruments used in the observations (\cref{tab:observational-data}).

\section{Model}
\label{sec:model}

The number of burst detections in each cycle is limited, and we therefore used a phase-folding method to model the periodic behaviour of FRBs. The phase of each burst was calculated using the equation
\begin{equation}
    \phi = \phi_{\text{ref}} + \frac{ \text{\textit{mod}}\del{P}}{P}\del{t - t_{\text{ref}}},
\end{equation}
where $t$ represents the time of the observation, $t_{\text{ref}}$ is the reference time for the reference phase $\phi_{\text{ref}}$, $P$ is the period of cyclic activity window, and \textit{mod} refers to a remainder operation that ensures all phases fall within the range of a single cycle.

For FRB~20121102A, we adopted a reference epoch of MJD \num{58356.5}, corresponding to $\phi_{\text{ref}} = 0$ as defined by \citet{Braga_2025}. Similarly, for FRB~20180916B, the reference epoch was taken to be MJD \num{58369.4} for $\phi_{\text{ref}} = 0$ following \citet{bethapudi2023high}.  

\begin{figure}[t]
    \centering
    \includegraphics[width=9cm]{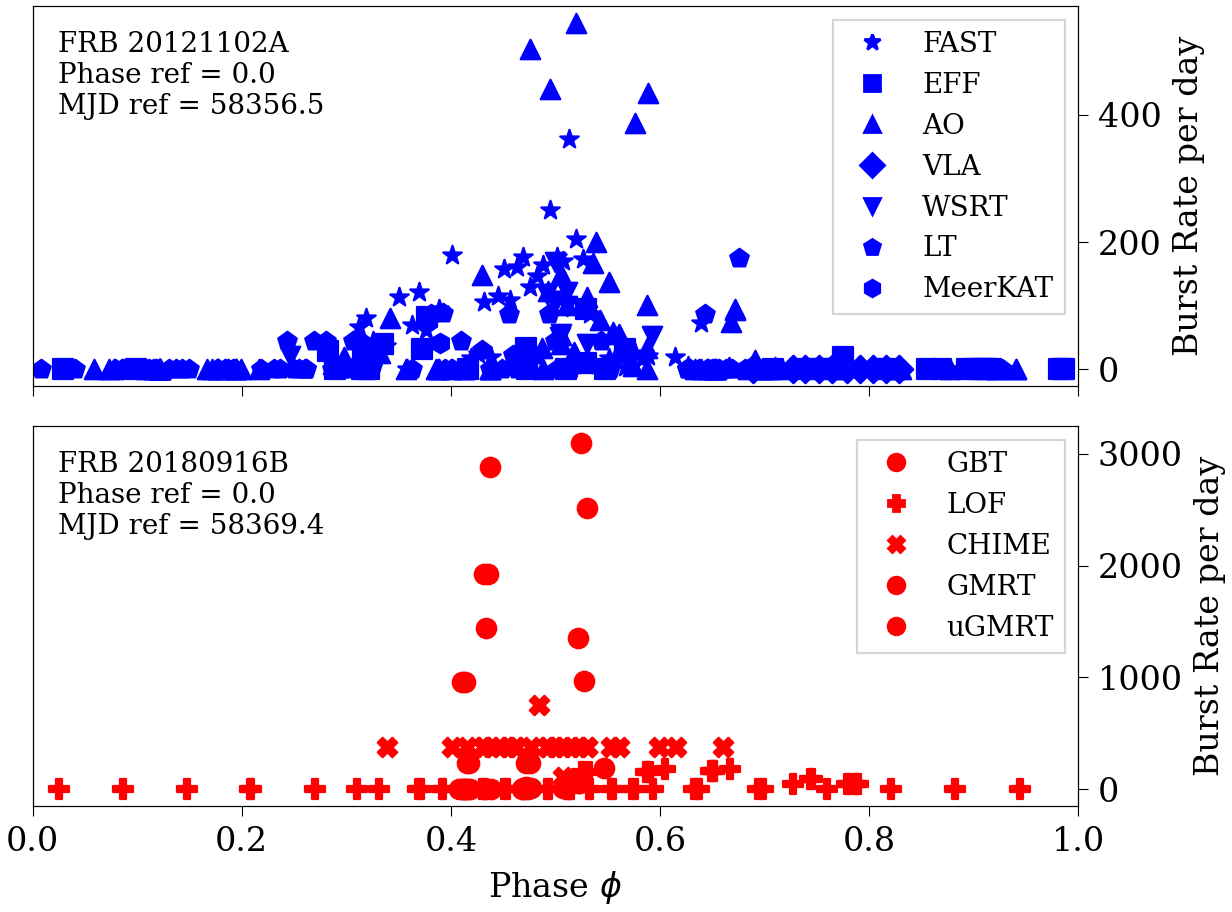}
    \caption{Folded burst-rate distribution for FRB~20121102A at L band (top)  and for FRB~20180916B at P band (bottom).
    The bursts are symbol-coded according to the observing radio telescope. The burst rate was scaled to a reference 100 m telescope using Eq. \ref{eq:rates}. A full list of the telescopes contributing to the L-band and P-band data can be found in Sect. \ref{sec:data_compi}. }
    \label{fig:gaussian-like}
\end{figure}

Taking all of the data described in Sect. \ref{sec:data_compi} and using the phase-folding algorithm, to a period of \num{159.3} days for FRB~20121102A and of \num{16.33} days for FRB~20180916B, we obtained Fig. \ref{fig:gaussian-like}. Fig. \ref{fig:gaussian-like} clearly shows burst detections from the two FRBs cluster near $\phi = 0.5$, resulting in a Gaussian-like activity window centred at mid-phase.

\subsection{Von Mises model}

Although the active window might indicate a Gaussian-like behaviour, this distribution is not suited for periodic data and would not allow for predictions of future activity cycles. To address this limitation, we adopted a Von Mises distribution, which represents a continuous probability distribution and works to sample circular objects, such as periodic data \citep{kitagawa2022mises}.  The Von Mises distribution is defined as \citep{mardia2009directional}
\begin{equation}
    f(x; \mu, \kappa) = \frac{1}{2\pi I_0(\kappa)}e^{\kappa \cos(x-\mu)},
\end{equation}
where \( \mu\) is the position at which the distribution is clustered, \( \kappa\) is a concentration parameter that determines the width and steepness of the curve, and \(I_0 (\kappa)\) is the modified Bessel function of order zero. This distribution has a dominion of $x \in [0,2\pi]$, and thus, a conversion was made to map the phase from $0$ to $1$.

\subsection{Chromatic model}

Building on the work of \cite{bethapudi2023high}, we defined the chromatic activity model using a power-law relation that linked the observing frequency and phase of detected bursts to construct the activity window. The model is described by the following power-law ($f(\nu)$) equation:
\begin{equation}
  f(\nu) = B \times \left(\frac{\nu}{\nu_0}\right)^A,
  \label{power_law}
\end{equation}
where $\nu$ is the observed frequency, and $\nu_0$ is the reference frequency at which the periodicity model was originally defined. For FRB~20121102A, $\nu_0 = \SI{1360}{\mega\hertz}$, and for FRB~20180916B, $\nu_0 = \SI{600}{\mega\hertz}$, which both correspond to the centre of the frequency band at which most of the detections are made for each source. The parameters $A$ and $B$ were fitted from the data and indicate the steepness and amplitude of the frequency dependence, respectively.

Importantly, each detection was considered to extend over the full bandwidth of the receiver used in the observation. To account for differences in frequency range and bandwidth among the telescopes in our sample within a given observing band (P, L, S, or C band), we replicated each event every \SI{50}{\mega\hertz} until the full bandwidth of the corresponding instrument was covered. For example, when Effelsberg detected an event using its seven-beam receiver with a \SI{300}{\mega\hertz} bandwidth centred at \SI{1.4}{\giga\hertz}, six data points were added to the sample, each with central frequencies given by
\begin{equation}
    \nu_i = \nu_{\text{low}} + i \times \Delta \nu\quad \text{for } i = 0, 1, 2, \ldots, N - 1
\end{equation}
where $N$ is the number of replicated frequency points, calculated as $N = \floor{BW / \Delta \nu}$. We adopted a replication step of 
$\Delta \nu = \SI{50}{\mega\hertz}$, motivated by the typical bandwidth differences among the telescopes in the sample. This procedure provided a smoother frequency coverage than a simple split into the four bands (P, L, S, and C band). We acknowledge that using the observed frequency extent of each burst would be a better approach. Nevertheless, most works do not report this information, and performing an individual analysis for individual burst is beyond the scope of this work.

\begin{figure}[t]
    \centering
    \includegraphics[width=9cm]{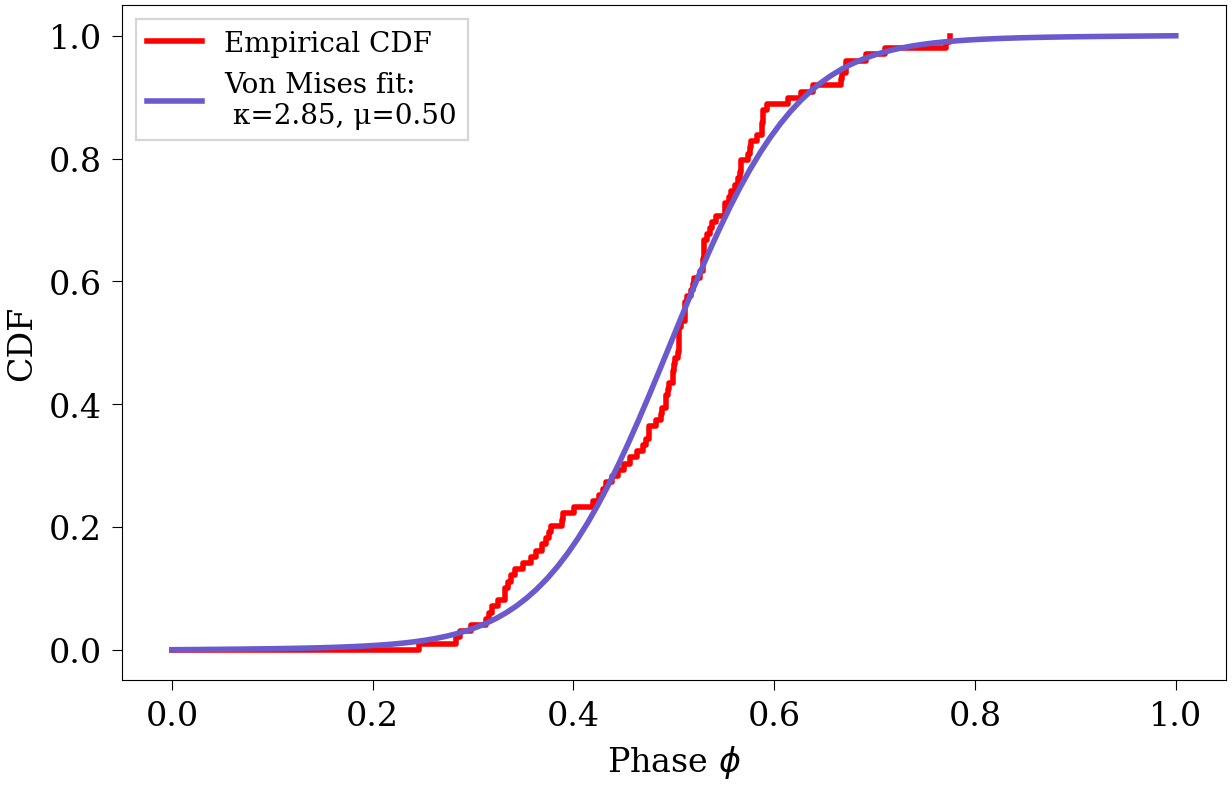}
    \caption{Von Mises CDF fit (purple) to the empirical CDF (red) of FRB~20121102A at \SI{1360}{\mega\hertz}. The fit is centred at a phase $\mu = 0.5$ and shaped by $\kappa = 2.8$ }
    \label{fig:CDF_1400_R1}
\end{figure}

\begin{figure}[t]
    \centering
    \includegraphics[width=9cm]{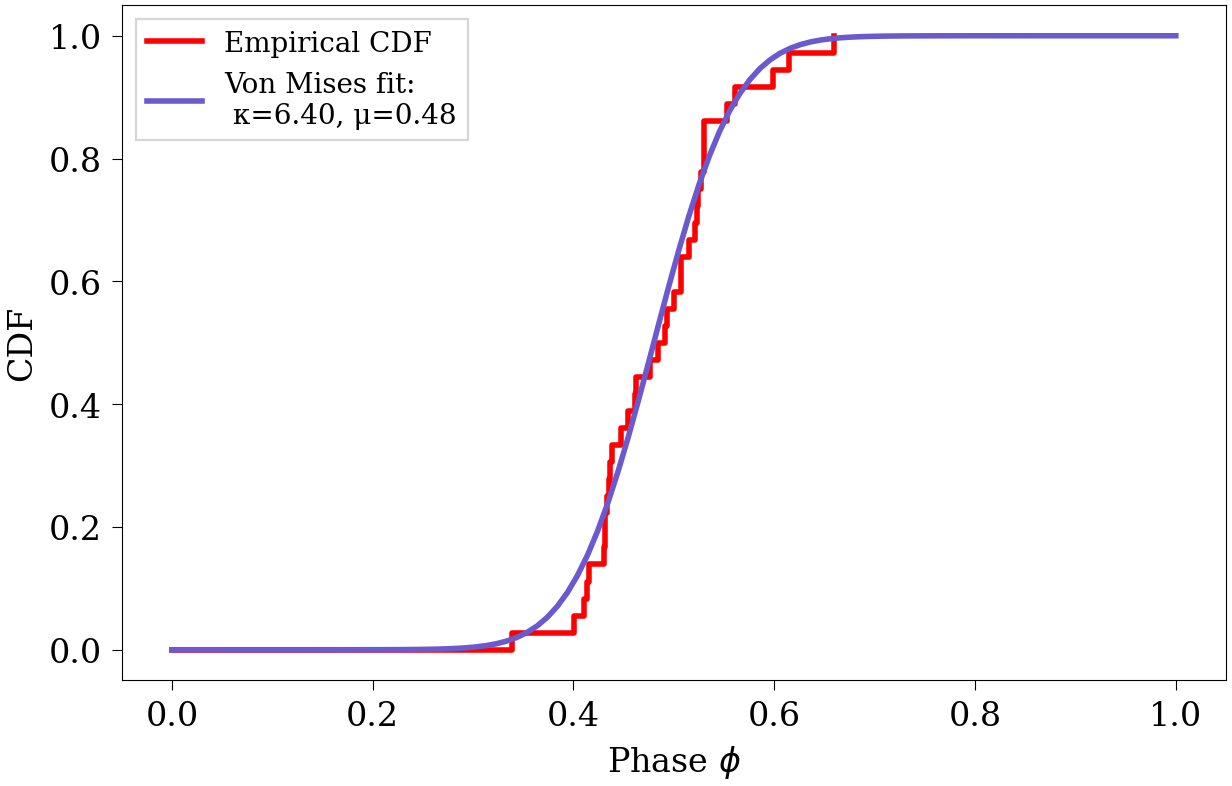}
    \caption{Von Misses CDF fit (purple) to the empirical CDF (red) of FRB~20180916B at \SI{600}{\mega\hertz}. The fit is centred at a phase $\mu = 0.48$ and shaped by $\kappa = 6.4$. }
    \label{fig:CDF_600_R3}
\end{figure}

The chromatic activity window was modelled by splitting the sample into frequency bins and fitting the cumulative distribution function (CDF) of a Von Mises to the phase distribution. The Von Mises distribution returns $\mu$, the peak of the distribution, and $\kappa$, the shape parameter that accounts for the steepness in the rise of the distribution. Examples of the fitting process are given in \cref{fig:CDF_1400_R1,fig:CDF_600_R3}, where we display the dataset at the reference frequency, 1.36\,GHz and 600\,MHz, at which most events are detected for FRBs 20121102A and 20180916B, respectively.

Then, the peak of the fit in each frequency bin, $\mu$, was fitted by a power law across frequency. This curve represents the peak activity of the window across frequencies. The spread around this peak was determined by the \textcolor{black}{phase} full width at half maximum (FWHM) of the Von Mises fit, which is related to the duration of the window. The activity window for a given frequency is therefore defined by
\begin{equation}
\label{eq:activity_window_phase}
    \phi_{\text{window}}(\nu) \in \left[ \mu(\nu) - \frac{\textcolor{black}{\delta(\nu)}}{2},\ \mu(\nu) + \frac{\textcolor{black}{\delta(\nu)}}{2} \right],
\end{equation}
with $\phi_{\text{window}}(\nu)$ the phase interval during which an FRB is expected to be active at a given observing frequency $\nu$, $\mu(\nu)$ the power law fitted through the peaks of the active windows across frequencies (i.e. $\mu$ of the Von Mises curve), and \textcolor{black}{$\delta(\nu)$ was obtained by fitting the power law to the FWHM measured in each frequency bin.}

\begin{figure}[t]
    \centering
    \includegraphics[width=9cm]{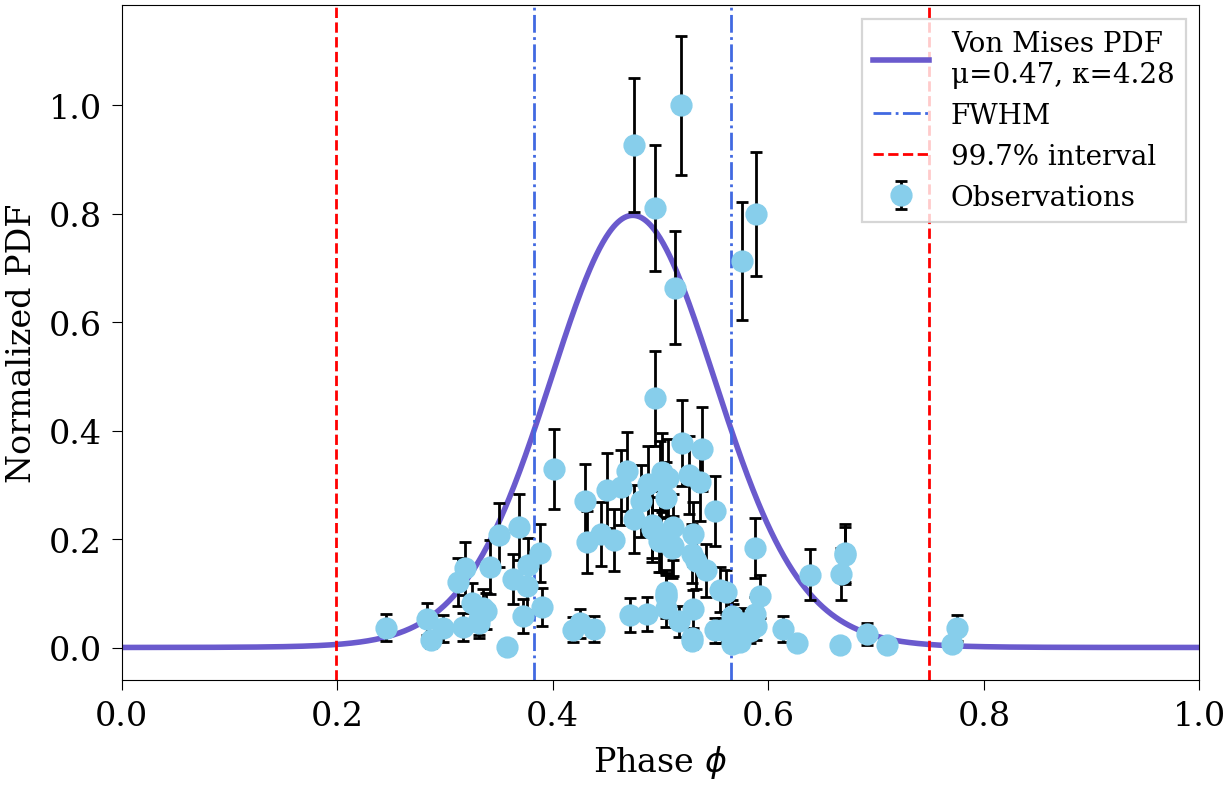}
    \caption{Modelled active window of FRB~20121102A at \SI{1360}{\mega\hertz}. The curve is centred at a phase $\mu = 0.47$ and shaped by $\kappa = 4.28$. The dashed-dotted blue limits indicate the FWHM, and the dashed red limits mark the activity window at $\SI{99.7}{\percent}$ confidence level. Observations are shown with blue points \textcolor{black}{with their respective $3\sigma$ uncertainties}}.
    \label{fig:CDF_active_window_1400_R1}
\end{figure}

\begin{figure}[t]
    \centering
    \includegraphics[width=9cm]{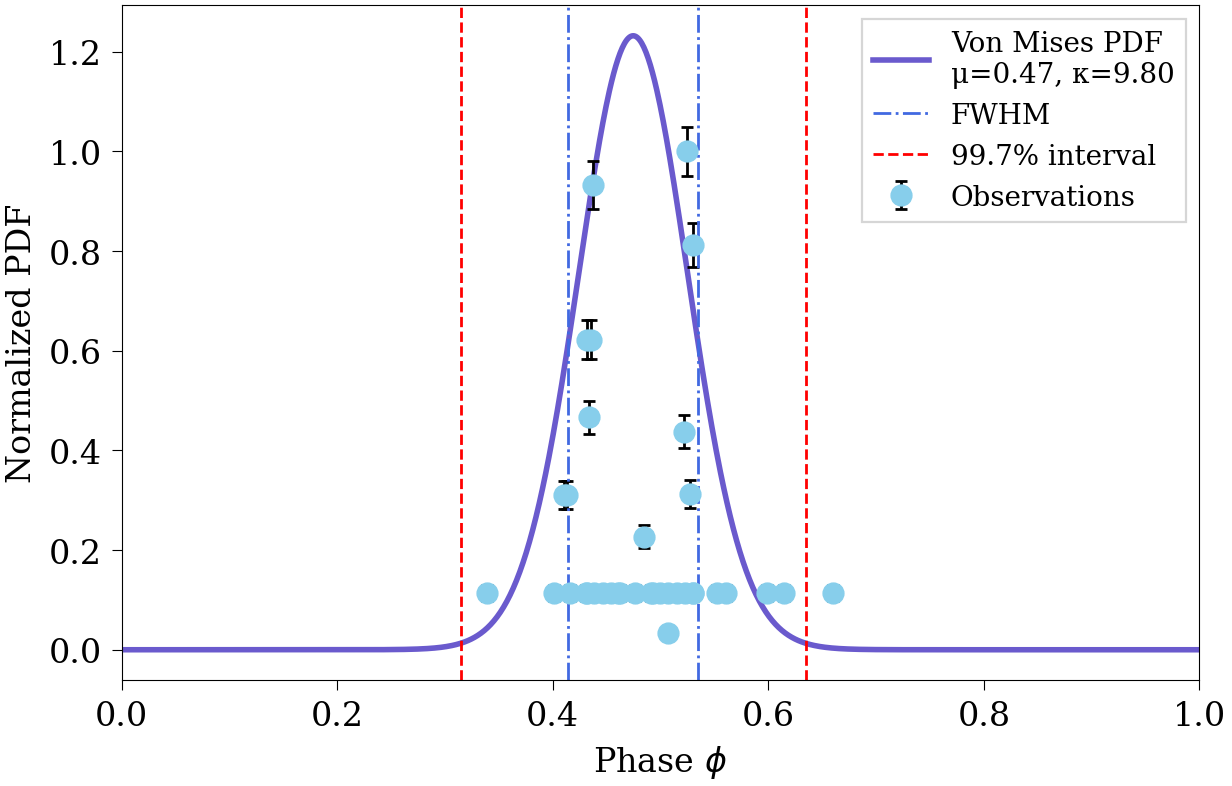}
    \caption{Modelled active window of FRB~20180916B at \SI{600}{\mega\hertz}. The curve is centred at a phase $\mu = 0.47$ and shaped by $\kappa = 9.8$. The dashed-dotted blue limits indicate the FWHM, and the dashed red limits mark the activity window at $\SI{99.7}{\percent}$ confidence level. Observations are shown with blue dots \textcolor{black}{with their respective $3\sigma$ uncertainties}}.
    \label{fig:CDF_active_window_600_R3}
\end{figure}

Figure \ref{fig:CDF_active_window_1400_R1} presents an example of the modelled active window of FRB~20121102A at \SI{1360}{\mega\hertz}, based on the chromatic model. The predicted Von Mises distribution is characterised by $\mu = 0.47$ and $\kappa = 4.28$. At a $\SI{99.7}{\percent}$ confidence level, and the width of the active window spans approximately \SI{55}{\percent} of the cycle ($\sim 87$ days). Similarly, for FRB~20180916B, Fig. \ref{fig:CDF_active_window_600_R3} shows the modelled active window at \SI{600}{\mega\hertz}, with a predicted Von Mises distribution with parameters $\mu = 0.47$ and $\kappa = 9.8$. The width of the window is estimated to be approximately \SI{32}{\percent} of the cycle ($\sim 5$ days) at the $\SI{99.7}{\percent}$ confidence level.

\section{Results}
\label{sec:results}

\begin{figure*}[t]
    
  \centering
  
  \subfloat[Chromatic activity window for FRB~20121102A.]{%
    \includegraphics[width=9cm]{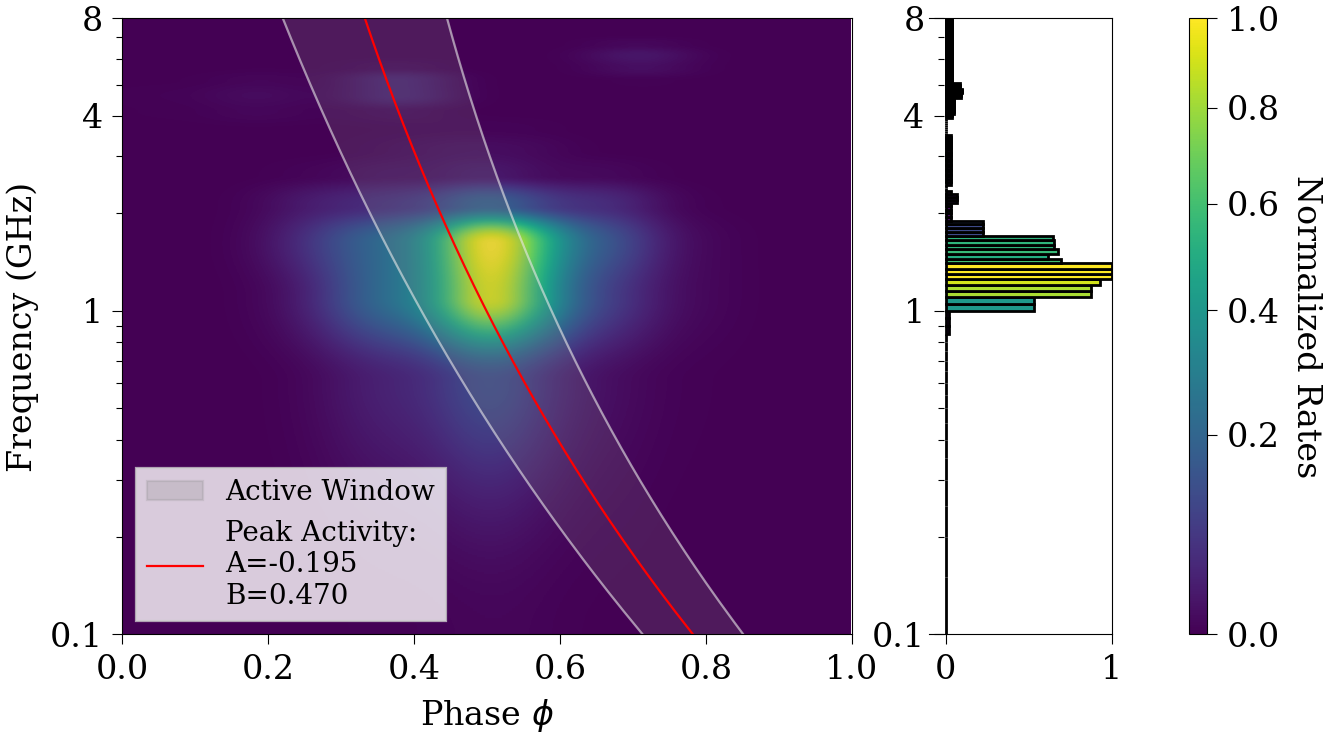}
    \label{fig:CDF_active_window_R1}
  }
  \hfill
  \subfloat[Chromatic activity window for FRB~20180916B]{%
    \includegraphics[width=9cm]{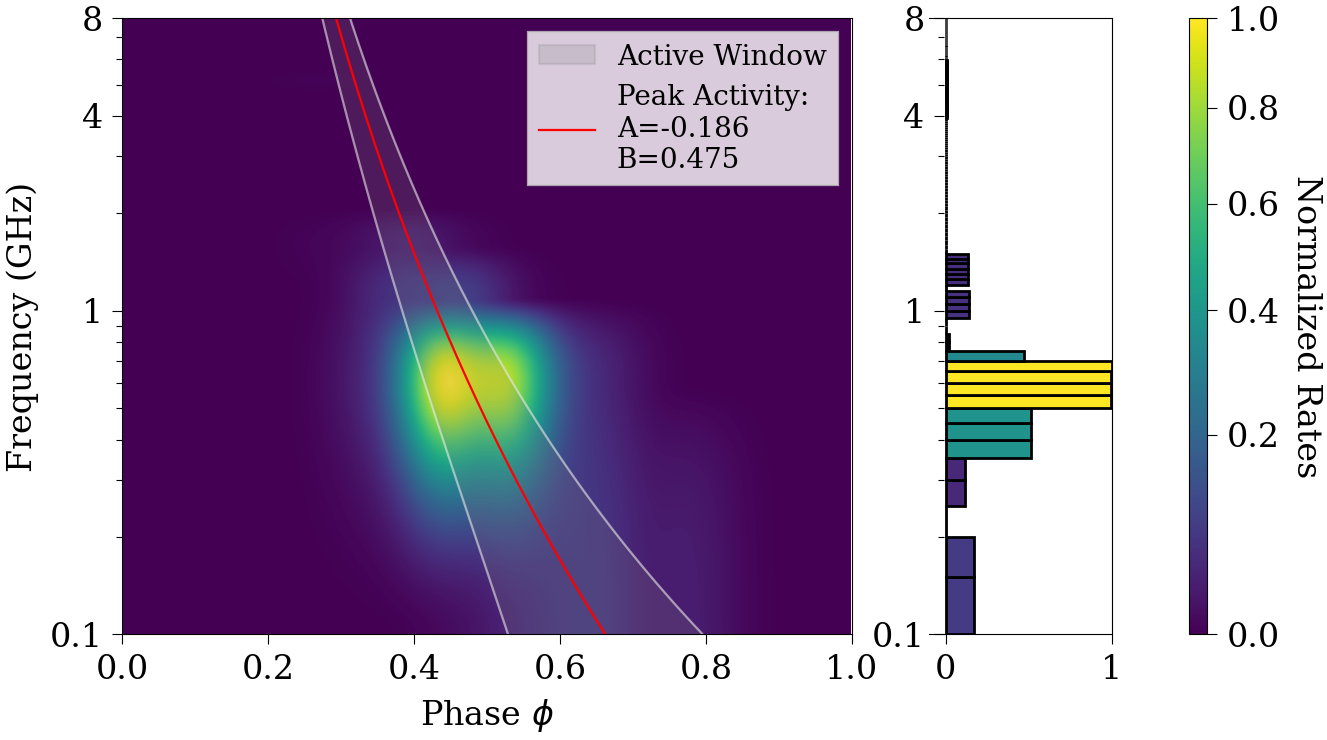}
    \label{fig:CDF_active_window_R3}
  }
  \caption{The activity window was constructed using a power-law model that fitted the peaks of the Von Mises distributions across frequency bands. The colour map shows the distribution of detected bursts using a Gaussian kernel density fitted to the data, included for visualisation purposes and normalised to resemble a probability density function.  The red curve corresponds to the peak of the activity window and the white limits show its FWHM. The marginal bins, colour-coded by density, display the distribution of burst detections across frequency.  For FRB~ 20121102A (panel a), the fitted parameters are $ A_{\mu} = -0.195$ and $ B_{\mu} = 0.470$, and for FRB~20180916B (panel b), the parameters are $ A_{\mu} = -0.186$ and $ B_{\mu} = 0.475$.}
  \label{fig:test}
\end{figure*}

\Cref{fig:CDF_active_window_R1,fig:CDF_active_window_R3} presents the chromatic activity windows of FRB~20121102A and FRB~20180916B. For FRB~20121102A, the denser detection region is in the L band, and for FRB~20180916B, the denser region is in the P band.

The frequency band for FRB~20121102A is split into bins \textcolor{black}{ \SI{50}{\mega\hertz} wide,} ranging from \SI{1}{\giga\hertz} to \SI{8}{\giga\hertz}. For FRB~20180916B, we used frequency bins of \SI{50}{\mega\hertz} in the range of \SI{0.1}{\giga\hertz} to \SI{5}{\giga\hertz}

The results of the power-law fitting for the parameters $A$ and $B$, as defined in Sect. \ref{sec:model}, are summarised in \cref{tab:fitting_power_law_CDF}. These parameters characterise the frequency dependence of the peak phase ($\mu$) and the FWHM (\textcolor{black}{$\delta$}) of the activity window, following the power-law relation presented in Sect. \ref{sec:model}.
Specifically, \( A_{\mu} \) and \( B_{\mu} \) refer to the fit of the peak phase, and \textcolor{black}{$A_{\delta}$} and \textcolor{black}{$B_{\delta}$} correspond to the FWHM fit.

\begin{table}[t]
\caption{\label{tab:fitting_power_law_CDF}
Best-fit power-law parameters for the activity window of FRBs.}
\centering
\begin{tabular}{lcc}
\hline\hline
\multicolumn{3}{c}{\textbf{FRB~20121102A}}\\
\hline
Parameter & $A_{\mu,\delta}$ & $B_{\mu,\delta}$ \\
\hline
Peak ($\mu$)        & $-0.195 \pm 0.073$ & $0.470 \pm 0.024$ \\
FWHM ($\delta$)     & $0.112 \pm 0.098$  & $0.185 \pm 0.014$ \\
\hline
\multicolumn{3}{c}{\textbf{FRB~20180916B}}\\
\hline
Parameter & $A_{\mu,\delta}$ & $B_{\mu,\delta}$ \\
\hline
Peak ($\mu$)        & $-0.186 \pm 0.018$ & $0.475 \pm 0.006$ \\
FWHM ($\delta$)     & $-0.444 \pm 0.047$ & $0.120 \pm 0.005$ \\
\hline
\end{tabular}

\tablefoot{
Best-fit parameters for the power-law model describing the frequency dependence of the activity window's peak phase ($\mu$) and FWHM (\textcolor{black}{$\delta$}) for FRB~20121102A and FRB~20180916B. The fit follows the relation presented in \cref{sec:model}. Parameter \( A \) characterizes the amplitude of the scaling relation, while \( B \) describes the steepness of the frequency dependence. Subscripts \( \mu \) and \textcolor{black}{\( \delta \)} are used to distinguish between the fits corresponding to the peak phase and the FWHM, respectively. The reported uncertainties represent \(1\sigma\) confidence intervals, derived from the square root of the diagonal elements of the covariance matrix obtained during the fitting process.}
\end{table}

\begin{figure*}[ht]
  \centering

  \subfloat[Predicted cycles of FRB~20121102A]{%
    \includegraphics[width=1\textwidth]{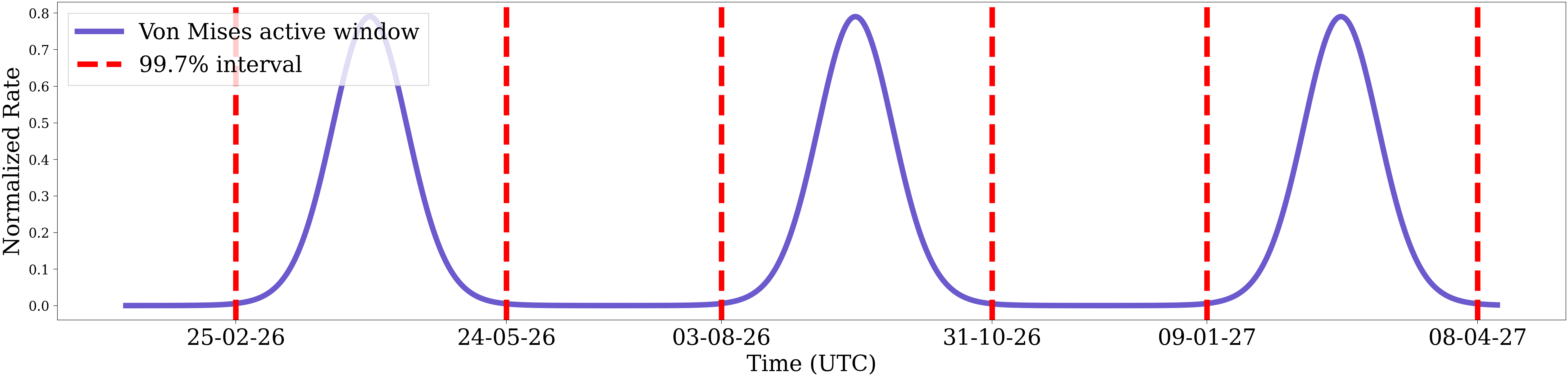}
    \label{fig:active_window_R1_pred}
  }

  \vspace{0.1cm} 

  \subfloat[Predicted cycles of FRB~20180916B]{%
    \includegraphics[width=1\textwidth]{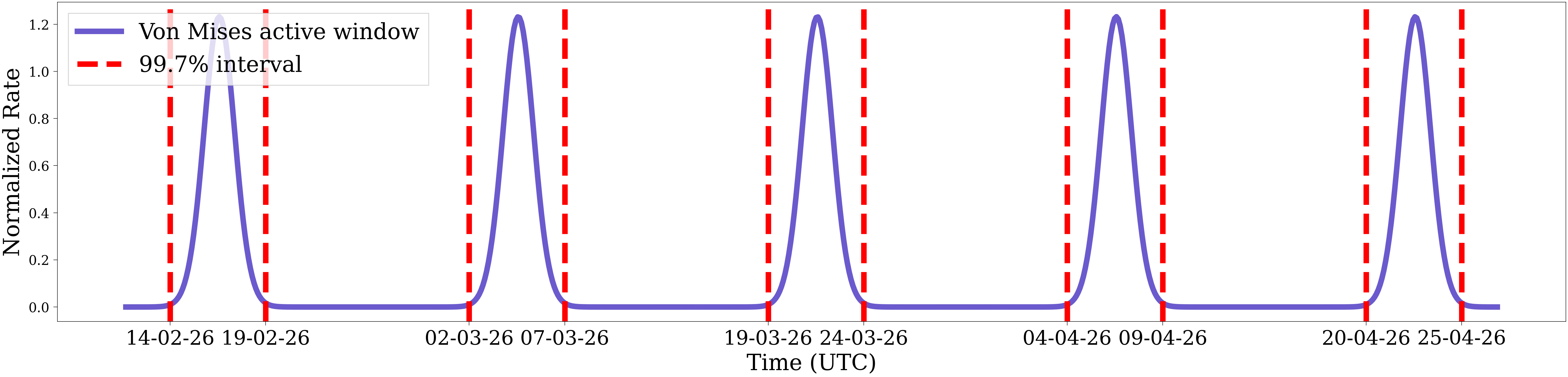}
    \label{fig:active_window_R3_pred}
  }

  \caption{Activity windows modelled using a Von Mises distribution. The dash-dotted red lines indicate the $\SI{99.7}{\percent}$ confidence level of the activity window. Panel (a): Predicted active windows of FRB~20121102A at \SI{1360}{\mega\hertz}, spanning from January \num{20}, 2026, to April \num{15}, 2027. Panel (b): Predicted active windows of FRB~20180916B at \SI{600}{\mega\hertz}, spanning from February \num{2}, 2026, to April \num{28}, 2026.}
  \label{fig:test}
\end{figure*}

For FRB~20121102A, \cref{fig:active_window_R1_pred} shows the predicted activity windows in UTC spanning from January \num{20}, 2026, to April \num{15}, 2027, computed at \SI{1360}{\mega\hertz}. We expect the next active window to begin around February \num{25}, 2026, and to end around May\num{24}. Likewise, \cref{fig:active_window_R3_pred} presents the predicted activity windows for FRB~20180916B at \SI{600}{\mega\hertz}, spanning February \num{2}, 2026, to April \num{28}, 2026. The upcoming active cycle is expected to start on February \num{14}, 2026, and to conclude on February \num{19}, 2026.

\section{Discussion}
\label{sec:disc}

\subsection{Active window behaviour}

For FRB~20180916B, the activity window starts earlier at a higher frequency. This is shown by the best-fit parameters for the peak of the active window, which are \( A_{\mu} = \num{-0.186 \pm 0.018} \) and \( B_{\mu} = \num{0.475 \pm 0.006} \), and indicate a clear chromatic dependence.
This behaviour was previously reported by \citet{pastor2021chromatic} and \citet{bethapudi2023high}, and our results are consistent with theirs while incorporating a larger dataset than was used in these previous studies. Regarding the duration of the activity window, which is defined by the FWHM (\textcolor{black}{\( \delta \)}), there is strong evidence of a narrowing with increasing frequency, indicated by the best-fit parameters (\( \textcolor{black}{A_{\delta}} = \num{-0.444 \pm 0.047} \) and \( \textcolor{black}{B_{\delta}} = \num{0.120 \pm 0.005} \)).

The comparison to the results of \citet{bethapudi2023high} for the peak activity (\( \mu \)) shows that our power-law fit yields \( A_{\mu} = \num{-0.186 \pm 0.018} \) and \( B_{\mu} = 0.475 \pm 0.006 \). These results agree with those reported by \citet{bethapudi2023high}, who found \( A_{\mu} = -0.23 \pm 0.05 \) and \( B_{\mu} = 0.47 \pm 0.02 \). Both results reveal a statistically significant chromatic trend in the activity peak, with a negative value of \( A_{\mu} \) that indicates that the start of the activity begins earlier at higher frequencies. Although our inferred chromaticity is less prominent for FRB~20180916B, it remains statistically consistent with the value reported by \citet{bethapudi2023high} within \( \sim 1\sigma\).

In the case of the FWHM, we converted our fitted \( B_{\delta} \) parameter to hours using the known FRB~20180916B period. We obtained \( \textcolor{black}{A_{\delta}} = -0.444 \pm 0.047 \) and \( \textcolor{black}{B_{\delta}} = \SI{46 \pm 1}{\hour} \).  \citet{bethapudi2023high} reported values of \( \textcolor{black}{A_{\delta}} = -0.35 \pm 0.32 \) and \( \textcolor{black}{B_{\delta}} = \SI{53.4 \pm 15.6}{\hour} \). In our results, the width of the activity window depends significantly on frequency, with a negative slope suggesting a narrower window at higher frequencies.  \citet{bethapudi2023high} reported a lesser negative slope, \textcolor{black}{meaning that in comparison with our modelling, their activity is wider at each frequency band}, but with a high uncertainty that is statistically consistent with zero. We report lower uncertainties because our sample is larger and because we used the CDF approach, which strengthens the observed behaviour trend. In terms of the amplitude parameter \textcolor{black}{$B_{\delta}$}, our results agree within $1\sigma$.

Regarding FRB~20121102A, our result also suggests a chromatic active window, with the activity starting  earlier at a higher frequency (\( A_{\mu} = \num{-0.195 \pm 0.073} \) and \( B_{\mu} = 0.470 \pm 0.024 \)). In contrast to FRB~20180916B, the FWHM \textcolor{black}{$\delta$} of FRB~20121102A widens with increasing frequency ($\textcolor{black}{A_{\delta}} = 0.112 \pm 0.098$ and $\textcolor{black}{B_{\delta}} = 0.185 \pm 0.0014$).

\subsection{FRB comparison}

\begin{figure}[t]
    \centering
    \includegraphics[width=\linewidth]{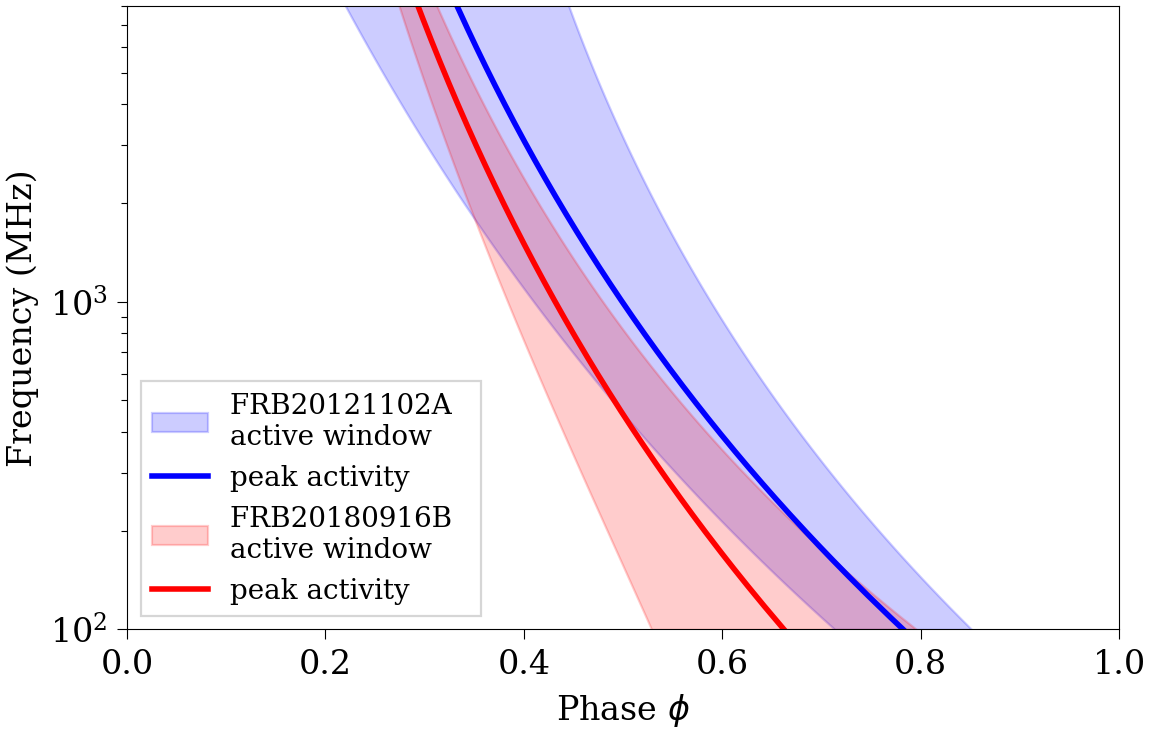}
    \caption{Comparison of the chromatic activity windows for FRB~20121102A (red) and FRB~20180916B (blue). The solid lines represent the power-law fits to the peak phases of each FRB activity window. For FRB~20121102A, the fit is characterised by $A_{\mu} = \num{-0.195 \pm 0.073} $ and $B_{\mu} = \num{0.470 \pm 0.024}$ (red line), and for FRB~20180916B, the fit is given by $A_{\mu} = \num{-0.186 \pm 0.018}$ and $B_{\mu} = \num{0.475 \pm 0.006}$ (blue line). The shaded regions represent the FWHM of the active windows.} 
    \label{fig:comparison_CDF}
\end{figure}

The results of the power-law fitting presented in \cref{tab:fitting_power_law_CDF} allowed us to compare the chromatic behaviour of FRB~20121102A and FRB~20180916B in terms of the peak phase and the FWHM of the active window, as shown in Fig. \ref{fig:comparison_CDF}.

The peak activity is similar for the two sources. It has a negative exponent \( A_{\mu} \), which indicates a shift in the activity window towards earlier phases at higher frequencies.
FRB~20121102A displays a slightly steeper frequency dependence in its peak phase, with \( A_{\mu} = \num{-0.195} \pm \num{0.073} \), compared to FRB~20180916B, for which \( A_{\mu} = \num{-0.186} \pm \num{0.018} \), but they agree within their uncertainties. The $B_{\mu}$ values are nearly identical for the two sources ($B_{\mu} \sim 0.47$), suggesting a similar scaling of the peaks with frequency.

In contrast, the \textcolor{black}{chromatic} behaviour of the FWHM differs between the two sources. For FRB~20121102A, the FWHM exhibits a broadening with increasing frequency, with a positive exponent $\textcolor{black}{A_{\delta}}$. FRB~20180916B shows a narrowing trend, with a negative exponent. The difference in the $\textcolor{black}{B_{\delta}}$ exponents indicates that the frequency scaling of the width is stronger for FRB~20121102A than for FRB~20180916B.

These results suggest that while the two sources show a similar frequency dependence in their peak phases overall, that is, they start earlier at higher frequencies, their \textcolor{black}{on-phase} behaviours diverge: The active phase of FRB~20121102A broadens at higher frequencies, whereas FRB~20180916B becomes narrower.

\subsection{Model discussion}

\begin{figure*}[t]
    
  \centering
  
  \subfloat[Scattered data of FRB~20121102A.]{%
    \includegraphics[width=9cm]{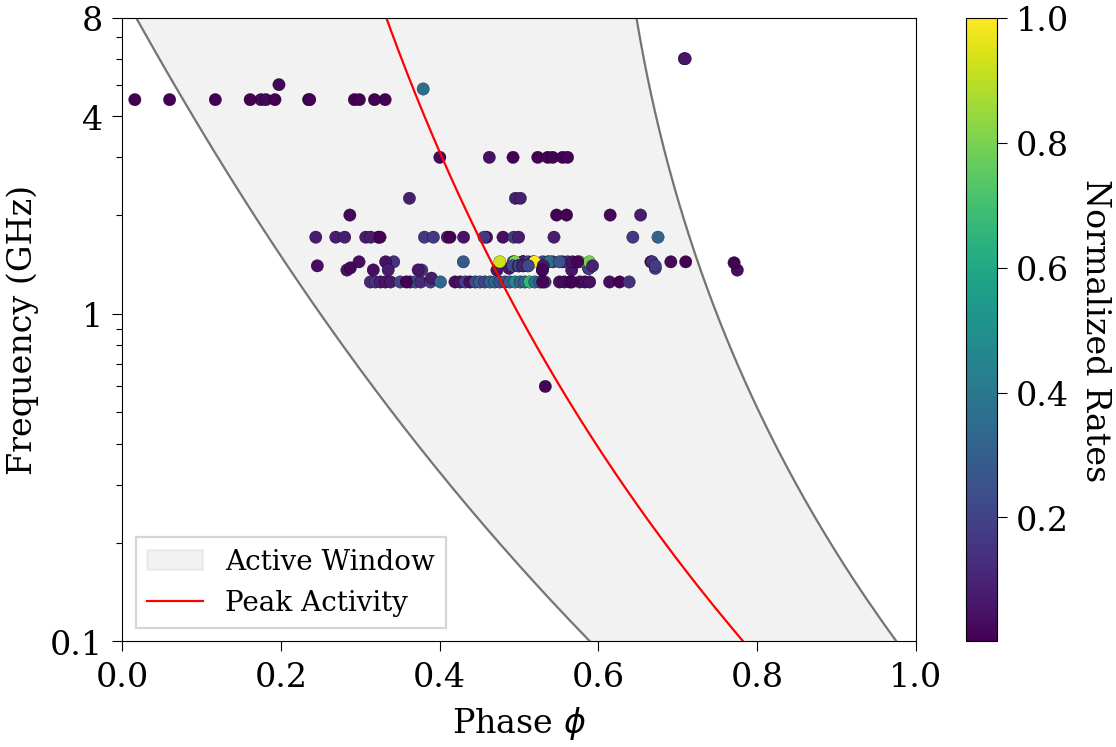}
    \label{fig:sc_R1_CDF}
  }
  \hfill
  \subfloat[Scattered data of FRB~20180916B]{%
    \includegraphics[width=9cm]{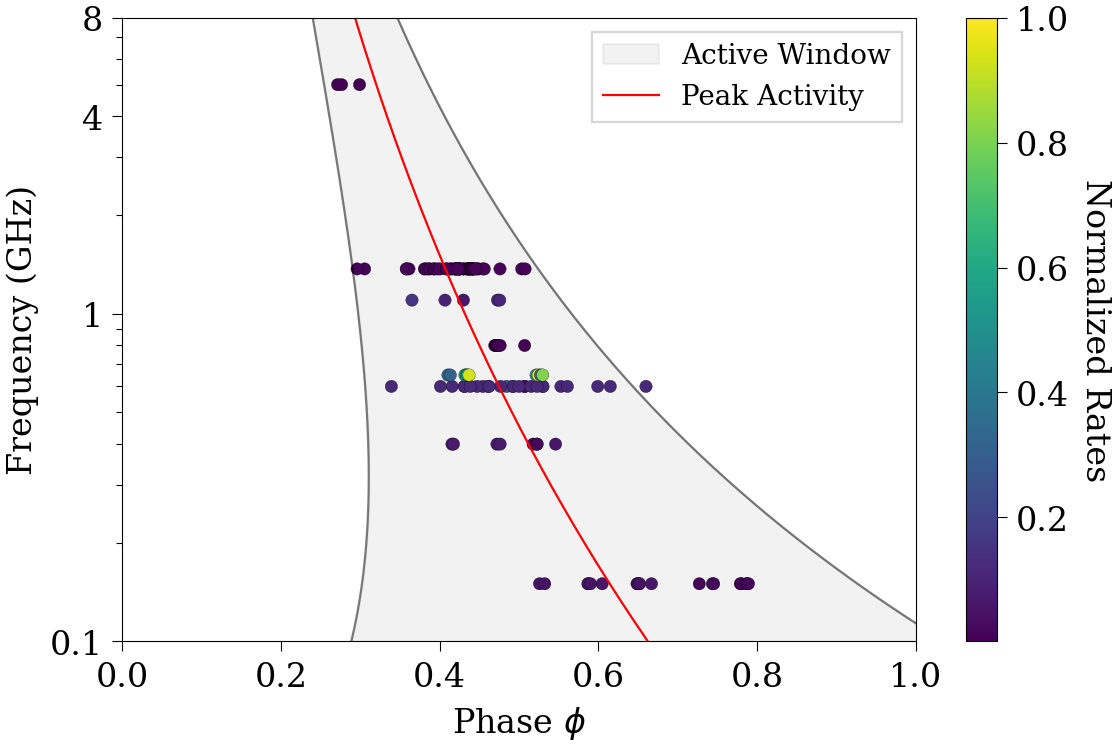}
    \label{fig:sc_R3_CDF}
  }
  \caption{Detections of FRB~20121102A (panel a) and FRB~20180916B (panel b) as scatter points. The colour scale depicts the normalised detection rate, with yellow indicating higher rates or densities. The active window is shown within the $\SI{99.7}{\percent}$ confidence level.}
  \label{fig:c_band}
\end{figure*}

The chromatic model, although simple in nature, allowed us to compare the differences in the activity for FRB~20180916B and FRB~20121102A. Possible model improvements might consider the rate scaling (Sect. \ref{sec:data_compi}) using the  $\gamma$-values reported by each telescope and using the width of the burst dynamic spectra instead of the observing bandwidth as a proxy.

It is important to note that the Von Mises CDF fit does not include information about the expected burst rates, but only about the phase distribution, which limits the amplitude of the curve, as shown in \cref{fig:CDF_active_window_1400_R1,fig:CDF_active_window_600_R3}. Any analysis related to the detection probability should take this limitation into account.

Bands with a small number of detections, such as C (4-8\,GHz), are particularly biased in the use of the observing bandwidth as an approximation for the frequency extent of the bursts, which complicates the fitting process because the observing setups at this frequency provide a few \si{\giga\hertz} wide band in comparison with fractions of \si{\giga\hertz} at lower frequencies. The C-band detections for the two FRBs are overlaid on the detection contour plot in Fig. \ref{fig:c_band}. In the case of FRB~20121102A, the C band shows that the detections cover most of the active window ($\sim \num{0.8}$ of the cycle) and that it starts earlier. This strengthens the findings up to S band of the FRB~20121102A chromaticity. Still, it is important to highlight that the full phase is not mapped.
In the case of FRB~20180916B, the detections in C band are extremely clustered in phase, which also impedes a proper fit. The limitations imposed by C band are not new. \cite{Braga_2025} also reported the limiting lack of observations for periodicity searches of FRB~20121102A over this frequency band. Further observations across a broader range of frequencies are necessary to fully address these limitations and provide a better understanding of the correlation between phase and frequency of the active window.

\textcolor{black}{We used the Kolmogorov--Smirnov (KS) goodness-of-fit test to evaluate whether the observed burst phase distributions were consistent with a Von Mises model. The KS statistic is defined as the maximum difference between the empirical cumulative distribution function (CDF) of the data and the CDF of the proposed model. In this context, the null hypothesis $H_0$ states that the burst phases are drawn from a Von Mises distribution, while the alternative hypothesis $H_1$ states that the observed phase distribution deviates from this model.  In particular, $p < 2.7\times10^{-3}$ ($3\sigma$ threshold) rejects the null hypothesis. }

\begin{figure}[t]
    \centering
    \includegraphics[width=\linewidth]{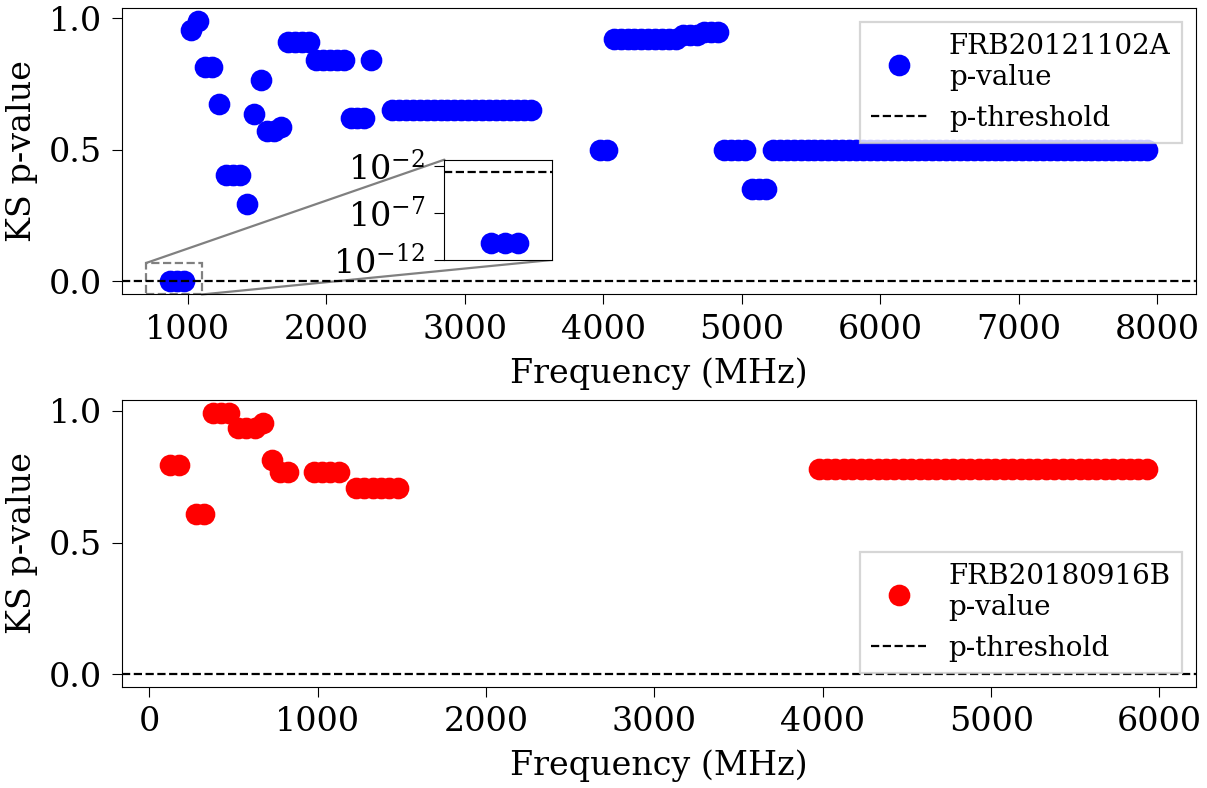}
    \caption{$p$ values of KS goodness-of-fit tests comparing Von Mises cumulative distribution function (CDF) models to the empirical phase CDFs, shown for each frequency bin of FRB~20121102A (top) and FRB~20180916B (bottom). For each frequency bin, the $p$ value quantifies whether the observed burst phase distribution is consistent with a Von Mises model. The dashed horizontal line marks the $3\sigma$ rejection threshold ($p = 2.7\times10^{-3}$); values below this line indicate that the Von Mises model is statistically rejected. The inset highlights the $p$ values in the lowest-frequency bins for FRB~20121102A.}
    \label{fig:chi_square}
\end{figure}

We applied the test to each Von Mises CDF fit at each frequency bin and present the results in Fig. \ref{fig:chi_square}. \textcolor{black}{For FRB~20180916B, the $p$ values across all frequency bins show no evidence of deviations from the model. For FRB~20121102A, the $p$ values indicate consistency with the model in most frequency bins, with $128$ out of $131$ frequencies having $p$ values above the commonly used reference levels}. In the range of lower frequencies, $\sim\SI{800}{\mega\hertz}$ to \SI{900}{\mega\hertz}, however, it is not able to characterise the data because the sampling size is too small. \textcolor{black}{We also note that at the highest frequencies, above \SI{4}{\giga\hertz} for FRB~20180916B and above \SI{5}{\giga\hertz} for FRB~20121102A, the $p$ values may be biased, again because of the small sample sizes in these bins}.

Although scaled rates depend on the $\gamma$ index value, we observed no significant change in the power-law fitting results when we tested the range of $\gamma=-1.1$ to $-1.8$ for FRB~20121102 or the $\gamma$ range of $-0.5$ to $-1.4$ for FRB~20180916B.

\subsection{Model implications}
The finding of chromaticity in FRB~20180916B imposed a constraint on progenitors models; models had to explain activity that started earlier and narrowed at higher frequencies. \citet{Li_2021} proposed that the chromatic behaviour of FRB~20180916B is consistent with theories where the periodicity is given by a slowly rotating or freely precessing magnetar. This mechanism is based on radius-to-frequency mapping (RFM) or altitude-dependent emission. The emissions at lower frequency originate at a higher altitude in the magnetosphere, where the lines of magnetic field diverge, creating an wider emission cone. The chromatic behaviour would arise from the emission cone staying aligned with the observer for longer, and the phase drift would be due to the higher magnetic polar angles observed at lower frequency bands.

\textcolor{black}{\citet{2019ApJ...879....4W} proposed ultra-long period magnetars (ULPMs) as an explanation for repeating FRBs, where stress accumulates in a low-twist magnetosphere and is released during short magnetar bursts, producing coherent radio emission that escapes as FRBs. Building on this framework, \citet{Bilous_2025} applied the ULPM model to FRB~20201124A}, which resembles the chromatic behaviour of FRB~20180916B. In the model, the RFM is used to explain the frequency dependence of the active window, but the authors noted a large difference in timescales for which the model must account. While pulsar active windows last fractions of a second, FRB activity windows can last for days. The authors argued that this discrepancy can be naturally resolved if FRBs originate from ULPMs with spin periods of several weeks or longer.

In addition to the long periodicity, FRB~20121102A introduces a new challenge; the broadening of the activity window at higher frequencies. \citet{Lyutikov_2020} proposed a binary comb model (BCM) that is generated when a pulsar orbits a massive O/B star and interacts with its wind. The periodicity arises from the free-free absorption in the primary wind of the star. The stellar wind is optically thick and too dense for radio waves, and a transparency funnel in the stellar wind is thus created by the pulsar. The bursts would only be visible when the line of sight is aligned with this cone. Since the free-free strongly depends on frequency, the wind will become more transparent at higher frequencies, which would allow bursts to escape from a wider range of angles. This would create an extended active window at a higher frequency. 

\citet{Ioka_2020} included two additional opacity models for the BCM. The fist model is the $\tau$-crossing mode, where the orbit intersects the photosphere, the pulsar FRB enters and exits the photosphere periodically, and the periodicity is explained by the temporal variation of the optical depth. The second model is the inverse funnel mode, in which the companion wind plunges into the wind of the pulsar FRB, inverting the direction of the funnel. In both of these cases, the implication is that the active window is wider at higher frequencies.

While FRB~20180916B shows a narrowing activity window at higher frequencies, seemingly contradicting the BCM predictions, FRB~20121102A presents a chromatic behaviour that is consistent with the model. Progenitor models must reconcile these differences within a single framework. Importantly, it should be differentiated whether the window dependence with frequency is intrinsic to the emission, due to propagation, or due to an observational bias \citet{Wada_2021}.

\subsection{Isotropic energy}

\begin{figure*}
    \centering
    \includegraphics[width=0.8\linewidth]{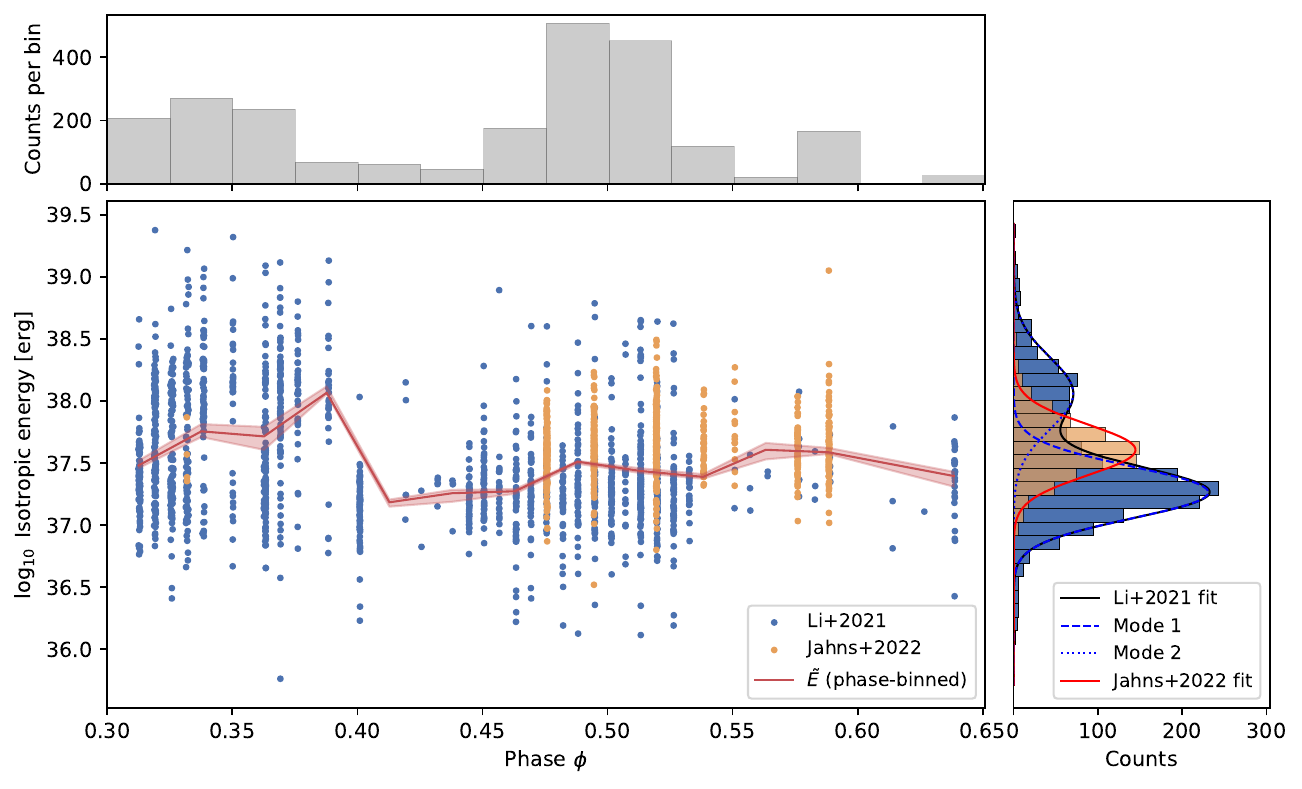}
    \caption{Phase–resolved burst energetics of FRB~20121102A. The central panel shows the isotropic burst energy as a function of activity phase for the \citet{2021Natur.598..267L} and \citet{jahns2022frb} samples. The solid red curve traces the phase-binned median energy, with the shaded region indicating the $1\sigma$ bootstrap uncertainty based on 10,000 resamples. The top panel shows the total number of bursts per phase bin. The burst phases were computed assuming a 159.3-day activity period, with a reference epoch MJD${\mathrm{ref}} = 58356.5$ corresponding to $\phi{\mathrm{ref}} = 0$ \citep{Braga_2025}. The right panel displays the marginal energy distributions for the two samples; the \citet{2021Natur.598..267L} sample is described by a bimodal Gaussian with peaks at $E \simeq 1.8\times10^{37}$ (dashed blue curve) and $1.2\times10^{38}$~erg (dotted blue curve), while the \citet{jahns2022frb} sample is consistent with a single Gaussian peaking at $E \simeq 4.0\times10^{37}$~erg (red curve).}
    \label{fig:new _energy}
\end{figure*}

With the dataset, we also corroborated whether the burst properties exhibit a phase-dependent behaviour. We analysed the distribution of isotropic burst energies, the waiting times, and the burst width for FRB~20121102A as a function of activity phase.

We considered a subset of the data described in Sect. \ref{sec:data_compi}, specifically, from \citet{jahns2022frb} and \citet{2021Natur.598..267L}, where sub-bursts were excluded for simplicity. The energy values in these two samples were originally computed using different methods: \citet{2021Natur.598..267L} used the central observing frequency, whereas \citet{jahns2022frb} scaled the burst fluences to account for signal components extending beyond the observing bandwidth. To ensure consistency across the dataset, we re-calculated the isotropic energy for all bursts using the expression
\begin{equation}
  E = 4\pi\times D_L^2\times \frac{F\times BW}{1+z} \times \SI{e-23}{\erg},
\end{equation}
where $D_L$ is the luminosity distance, $F$ is the fluence in \si{\jansky\s}, $BW$ the bandwidth in \si{\hertz}, and $z$ is the redshift. The ToA between bursts were calculated as the difference in arrival time of two bursts $wt_i = \text{ToA}_{i+1}-\text{ToA}_{i}$. The fluences were obtained from the datasets provided by the studies. In the case of \citet{2021Natur.598..267L}, the fluences were measured using the peak flux density and equivalent burst duration, and for \cite{jahns2022frb}, they were calculated using an SEFD dependent on frequency and zenith angle. 

We performed simple correlation tests for each parameter (E, ToA, and burst width) and the activity phase. In general, the correlation between parameters and phase is negligible, except for the isotropic energy shown in Fig. \ref{fig:new _energy}. \textcolor{black}{We tested the combined dataset and the \citet{2021Natur.598..267L}  and  \cite{jahns2022frb} samples independently, first using a circular distance metric between the burst phase and the centre of the activity window. This was motivated by the results for FRB~20180916B, whose bursts energy seems to peak in middle of the activity window
 \citep{bhattacharyya2024wideband}. In the \citet{2021Natur.598..267L} sample, we found a positive correlation between the isotropic energy and the distance to the window centre (Spearman $\rho = 0.278$, $p = 1.2 \times 10^{-30}$; Kendall $\tau = 0.178$, $p = 2.7 \times 10^{-27}$), indicating that higher-energy bursts preferentially occur away from the centre. A weaker but still significant trend was observed for the combined dataset (Spearman $\rho = 0.141$, $p = 5.4 \times 10^{-12}$), while no significant correlation was found in the \cite{jahns2022frb} sample. The difference probably arises because the \citet{2021Natur.598..267L} sample spans most of the activity window, but includes fewer bursts near the end, whereas the \citet{jahns2022frb} sample predominantly covers the later half of the window and exhibits a more dispersed phase distribution.}

The circular distance test does not distinguish between the beginning and end of the activity window, however, and we therefore further tested for a monotonic dependence of the energy on phase. We found a negative trend in the \citet{2021Natur.598..267L} sample (Spearman $\rho = -0.271$, $p = 3.5 \times 10^{-29}$), which implies that the burst energies are higher at earlier phases and decrease toward the end of the activity window. This trend is also supported by the phase-binned median energy shown in Fig. \ref{fig:new _energy} with the red line ($1\sigma$ bootstrap uncertainties), which peaks away from the centre of the activity window. Importantly, the phase-resolved burst rate (top panel) shows that the centre of the activity window contains the highest number of detected bursts, but corresponds to one of the lowest median energies. This demonstrates that the observed trend is not necessarily due to variations in burst rate.

These results indicate a mild preference for more energetic bursts occurring at the beginning of the activity window in FRB~20121102A overall, in contrast to FRB~20180916B, where the burst energy has been reported to peak near the centre of the activity cycle \citep{bhattacharyya2024wideband}.

\section{Conclusions}
\label{sec:conclusions}
We presented the modelling of the chromatic activity windows of two periodically repeating FRBs, FRB~20121102A and FRB~20180916B.  The chromaticity was modelled using a combination of a Von Mises distribution to model the phase and a power-law relation to model the frequency dependence, as described in Sect. \ref{sec:model}.  
To construct the models, we used publicly available data of the sources as listed in Sect. \ref{sec:data_compi}, and we divided these data by frequency bands. The detections in \cref{fig:CDF_active_window_R1,fig:CDF_active_window_R3} cover a range of \SIrange{.1}{8}{\giga\hertz}.  

The two FRBs display a shift in peak activity to earlier phases at higher frequencies, although their steepness differs, with FRB~20121102A displaying a slightly steeper dependence on frequency than FRB~20180916B, as shown in Fig. \ref{fig:comparison_CDF}. For FRB~20121102A, however, the widening in the activity window occurs at higher frequencies than for FRB~20180916B, which displays a broadening at lower frequencies. For FRB~20180916B, the chromatic behaviour agrees with the behaviour presented by \cite{bethapudi2023high}. For FRB~20121102A, this is a newly reported feature. 

The difference between these two periodic FRBs is also reflected in the isotropic energy distributions of their bursts. For FRB~20180916B, the median burst energy has been reported to peak near the centre of the activity window, where we found evidence for more energetic bursts from FRB~20121102A at the beginning of the activity window. An in-depth modelling of the observational biases and a more homogeneous sampling of the active window is required to robustly confirm these trends.

\begin{acknowledgements}
This research started during a summer internship conducted at the Max-Planck-Institut für Radioastronomie, M.C.E extends her gratitude to M. Kramer and his research group, in particular to L. Spitler and S. Bethapudi for useful discussions. The authors thank the anonymous referee for their valuable comments on the manuscript that helped improve its quality. M.C.E thanks I. Crosby for the feedback. M. C acknowledges support from Max Planck Society through the Max Planck Parter Group at UC and CIA250016 (CPS-RTC). T.C.\ gratefully acknowledges support by the ANID BASAL FB210003.

\end{acknowledgements}

\bibliographystyle{aa}
\bibliography{aanda.bib}

\end{document}